\documentclass[twocolumn,showpacs,preprintnumbers,amsmath,amssymb,superscriptaddress,longbibliography]{revtex4-1}

\usepackage{graphicx}
\usepackage{dcolumn}
\usepackage{bm}
\usepackage{color}
\usepackage{pifont}
\usepackage{natbib}
\usepackage{hyperref}
\hypersetup{
colorlinks = true,
urlcolor   = blue,
linkcolor  = blue,
citecolor  = blue
}

\begin{document}

\title{A two-state generalisation of the strong collision model}

\author{Ola~Kenji~Forslund}
\email{ola.forslund@physics.uu.se}
\affiliation{Physik-Institut, Universität Zürich, Winterthurerstrasse 190, CH-8057 Zürich, Switzerland}
\affiliation{Department of Physics and Astronomy, Uppsala University, Box 516, SE-75120 Uppsala, Sweden}

\date{\today}

\begin{abstract}
Muon spin relaxation is a powerful technique for probing static and dynamic local magnetic fields. The strong collision model, based on a Gaussian-Markovian process, is commonly used to account for dynamical effects. Yet, it remains limited in describing systems where the local field undergoes discrete state changes. To address this, I introduce a generalized two-state strong collision model that explicitly incorporates transitions between distinct local field environments during fluctuations. This extension allows for a more accurate representation of dynamical effects, particularly in systems where each collision alters the underlying static polarisation function. Analytical and numerical solutions are presented, and the model's applicability is demonstrated {\color{black}and discussed} across relevant physical systems -- including low-dimensional magnets, systems with dynamic disorder and ion and muon diffusion. These results offer an enhanced framework for interpreting data in complex materials and extend the method's reach to a broader class of dynamical phenomena in condensed matter physics.
\end{abstract}


\maketitle

\section{\label{sec:Intro}Introduction}
Muon spin rotation, relaxation, and resonance ($\mu^+$SR) experiments provide a direct means of probing static and dynamic local magnetic field environments \cite{Nagamine2003}. Owing to its versatility, $\mu^+$SR has found widespread application across various scientific disciplines \cite{hillier2022}. While much of the focus has been on magnetic systems—particularly in studies of magnetism and superconductivity \cite{Blundell2004}—the technique has also proven valuable in chemistry \cite{Walker1983}, biology \cite{Torikai2001}, and, increasingly, in the study of ion diffusion in battery materials \cite{Sugiyama2009}.

In a $\mu^+$SR experiment, a muon is implanted into a material, where it settles near an electrostatic potential minimum, often close to an anion and at a distance of roughly 1~\AA~\cite{Sulaiman1994, Moller2013}. The local magnetic field at the muon site is primarily composed of dipolar contributions from nuclear and electronic magnetic moments. Due to its gyromagnetic ratio ($\gamma_{\mu}$), the muon undergoes spin precession in the presence of a magnetic field. The parity-violating weak interaction that governs muon decay leads to an asymmetric positron emission \cite{Lee1956}, which is detected to extract information about the muon spin polarisation.

The resulting depolarisation reflects how the muon ensemble responds to the local magnetic environment and is typically analysed using predefined static polarisation functions \cite{Yaouanc2011, musrfit}. However, when dynamic effects are present, time-dependent fluctuations in the local field introduce additional complexity. To incorporate such dynamics, the strong collision model \cite{Kehr1978} has been widely adopted in $\mu^+$SR studies \cite{Kojima1995, Crook1997, Lago2005}. This model assumes a Gaussian-Markovian stochastic process and extends static depolarisation functions into the dynamic regime through numerical treatment. While it offers a general and effective framework for modelling many dynamical systems, it remains limited in scope—particularly in systems where each field fluctuation alters the underlying static field configuration.

\begin{figure}[ht]
  \begin{center}
    \includegraphics[keepaspectratio=true,width=75 mm]{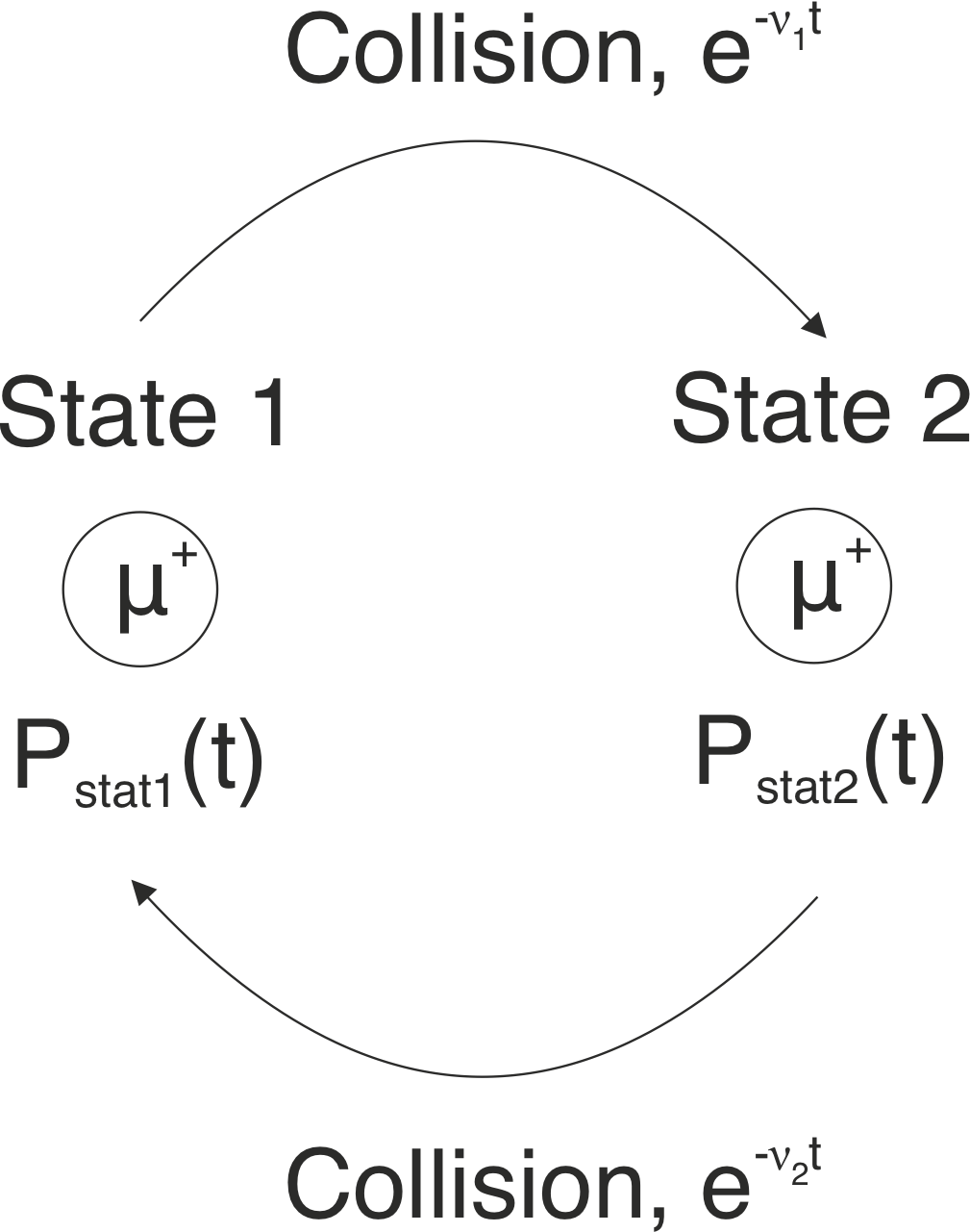}
  \end{center}
  \caption{Schematic overview of the two-state generalisation of the strong collision model. The local magnetic field environments, state 1 and state2, are each described by the static depolarisation functions $P_{\rm stat1}(t)$ and $P_{\rm stat2}(t)$, respectively. Transitions between the two states occur at each collision, governed by the probability $e^{-\nu_n t}$, where $\nu_n$ is the correlation frequency for transitions out of state $n$ (with $n = 1, 2$).
  }
  \label{fig:figure1}
\end{figure}

Motivated by these limitations, I introduce a two-state generalisation of the strong collision model. In contrast to the nominal strong collision model, which assume the static state remains unchanged between fluctuations, the two-state model allows for dynamic transitions between distinct local field environments during each stochastic event (Fig.~\ref{fig:figure1}). This more flexible formulation is particularly relevant to systems such as low-dimensional magnets, materials with dynamic disorder, and compounds exhibiting ion or muon diffusion. I present both analytical and numerical solutions and demonstrate the model’s relevance to key cases in condensed matter physics, where the conventional approach may fail to provide an accurate description.

\section{\label{sec:Intro}Strong collision model}
In the strong collision model, fluctuations in the local magnetic field are assumed to follow a Gaussian-Markovian process. Simply put, this means that the field takes on new, uncorrelated values drawn from a Gaussian distribution, and its evolution does not depend on its prior state. Under this assumption, the field autocorrelation function takes the form
\begin{eqnarray}
\langle B^{\alpha}_{\rm loc}(0)B^{\alpha}_{\rm loc}(t)\rangle =\langle(B^{\alpha}_{\rm loc})^2\rangle e^{-\nu t},
\label{eq:autoB}
\end{eqnarray}
where $\nu$ is the fluctuation (or correlation) frequency, $\alpha$ refers to Cartesian components, and the angle brackets denote an ensemble average over muon ensemble. In the strong collision model, the local magnetic field is assumed to remain static between collisions. At each collision event, the field abruptly changes to a new value that is statistically independent of its previous value. The time-dependent polarization function, $P_{\rm dyn}(t)$, is then constructed as the sum over contributions from muons experiencing different numbers of collisions between time 0 and $t$:
\begin{eqnarray}
P_{\rm dyn}(t) &=&
\sum_n^{\infty} R_n (t),
\label{eq:dyn}
\end{eqnarray}
where $R_n(t)$ represents the polarization function for muons that experience exactly $n$ collisions within the interval $[0,t]$. In essence, $P_{\rm dyn}(t)$ is composed of muons undergoing zero, one, two, and more collisions, each contributing to the overall depolarization profile. This problem has been extensively studied \cite{Karlsson1995, Yaouanc2011}, and can be expressed recursively as:
\begin{eqnarray}
R_{\rm n}(t) &=&
\nu \int_0^t R_{\rm n-1}(t_n)R_0(t-t_n)dt_n,
\label{eq:recur}
\end{eqnarray}
where $R_0(t)$ is the polarization function in the absence of any collisions (i.e., the static case). Here, $P_{\rm stat}(t)$ is the static polarization function and $R_0(t) = P_{\rm stat}(t) e^{-\nu t}$ reflects the effect of a single exponential survival probability. Although a general analytical solution in the time domain is not available, the Laplace transform offers a powerful method of solution. Using the property that $f_n(s) = \nu^n f_0(s)^{n+1}$ \cite{Yaouanc2011}, where $f_0(s)$ is the Laplace transform of $R_0(t)$, the Laplace transform of $P_{\rm dyn}(t)$ becomes a geometric sum:
\begin{eqnarray}
F(s)=L(P_{\rm dyn}) &=&
\sum_n \nu^n f_0(s)^{n+1}=\frac{f_0(s)}{1-\nu f_0(s)}.
\label{eq:Lap}
\end{eqnarray}
The strong collision model thus offers a convenient and widely-used framework for introducing dynamics into static polarization functions, particularly because the dynamic polarization is directly derived from its static counterpart. Its formulation relies fundamentally on the assumption of Markovian dynamics, in which the system has no memory of previous field values between collisions.

\begin{figure*}[ht]
  \begin{center}
    \includegraphics[keepaspectratio=true,width=\textwidth]{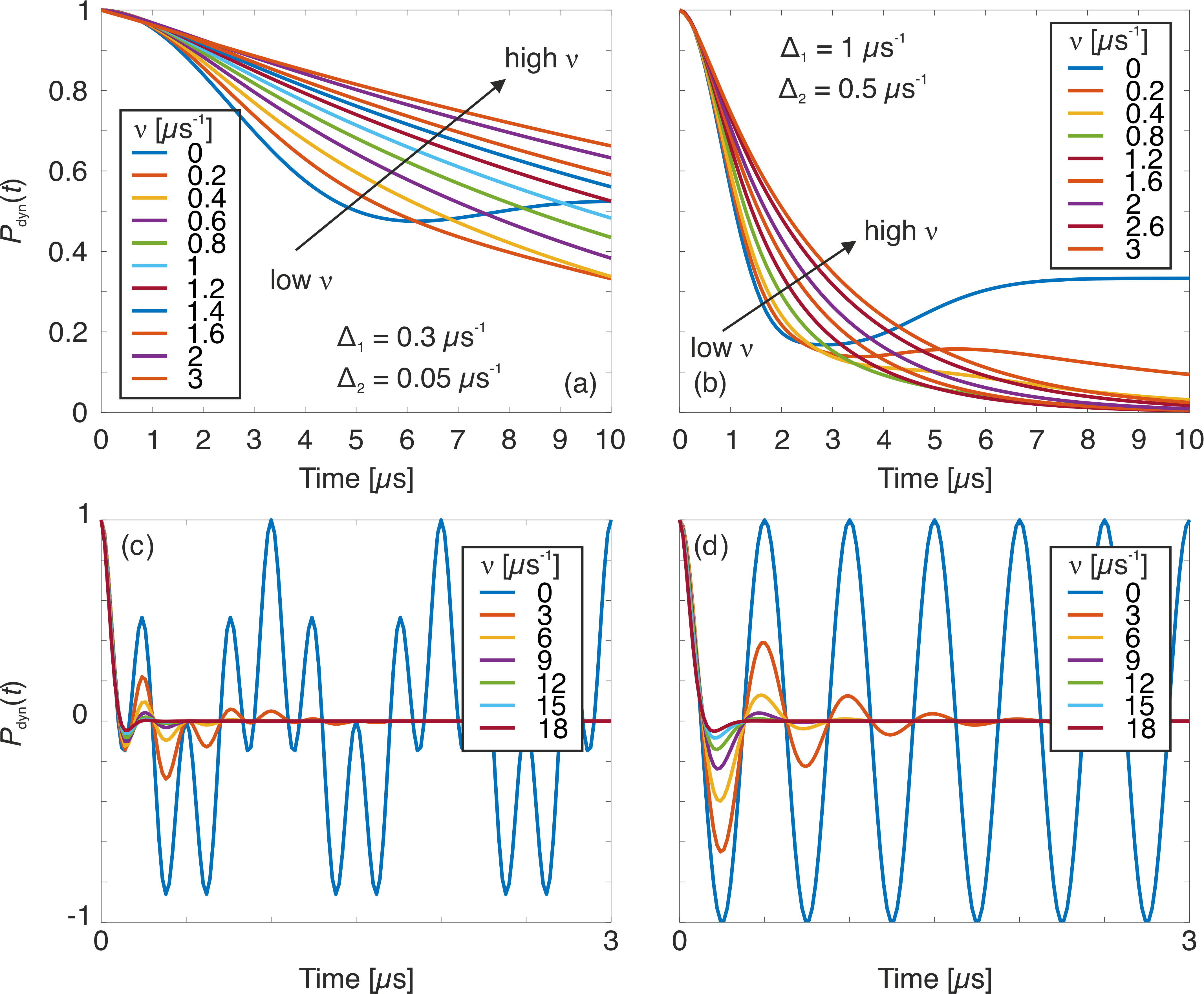}
  \end{center}
  \caption{(a, b) Solutions of Eq.~\ref{eq:recur2_3} for different values of $\Delta$, shown as a function of $\nu$, using a Gaussian Kubo-Toyabe function (Eq.~\ref{eq:kt}). In (a), $\Delta_1 = 0.3\mu$s$^{-1}$ and $\Delta_2 = 0.05~\mu$s$^{-1}$, while in (b), $\Delta_1 = 1~\mu$s$^{-1}$ and $\Delta_2 = 0.5~\mu$s$^{-1}$. (c, d) Solutions of Eq.~\ref{eq:recur2_3} as a function of $\nu$, using a cosine function (Eq.~\ref{eq:cos}). In (c), the frequencies {\color{black}($\gamma_\mu B$)} are 1 and 4~MHz, and in (d), they are 2 and -2~MHz. All figures use $g=0.5$ and $\nu=\nu_1=\nu_2$. {\color{black} $\nu$ is given in units of $\mu$s$^{-1}$ and can be compared to the characteristic field scales ($\Delta$ or $\gamma_\mu B$) to determine whether the system is in the quasi-static or motional narrowing regime, as explained in the main text.}
  }
  \label{fig:figure2}
\end{figure*} 

\section{\label{sec:Results}Two-state generalisation}
The nominal strong collision model assumes that the new state after each collision can be described by the same static polarisation function. However, it is reasonable to consider scenarios where the local field changes sufficiently due to the collision such that the new state is no longer described by the same static function. This situation is illustrated schematically in Fig.~\ref{fig:figure1}.

To account for such cases, I introduce a generalized two-state strong collision model in which the system alternates between two distinct local environments. Each environment is characterized by its own static polarisation function, denoted $P_{\rm stat1}(t)$ and $P_{\rm stat2}(t)$, corresponding to different local field configurations. Upon each collision, the system does not merely lose memory of its prior evolution but transitions into the complementary local field environment, described by the alternate static polarisation function. While the formalism retains the basic structure of the nominal model, it crucially incorporates switching between environments. This allows for a more realistic representation of systems where the local field itself evolves dynamically, rather than simply decorrelating.

To analytically solve this model, the contribution from muons undergoing different numbers of collisions must be considered. For example:
\begin{itemize}
    \item The contribution from zero collisions up to time $t$ is
    \begin{equation}
        R_0(t) = e^{-\nu t} P_{\rm stat1}(t).
    \end{equation}
    
    \item The contribution from one jump at time $t_1$ is
    \begin{equation}
        R_1(t) = \int_0^t e^{-\nu(t - t_1)} P_{\rm stat2}(t - t_1) \cdot e^{-\nu t_1} P_{\rm stat1}(t_1) \cdot \nu \, dt_1.
    \end{equation}
    
    \item For two jumps at $t_1$ and $t_2$, the contribution is
    \begin{multline}
        R_2(t) = \int_0^t \int_0^{t_2} e^{-\nu(t - t_2)} P_{\rm stat1}(t - t_2) \cdot e^{-\nu(t_2 - t_1)} \\
        \times P_{\rm stat2}(t_2 - t_1) \cdot e^{-\nu t_1} P_{\rm stat1}(t_1) \cdot \nu^2 \, dt_1 \, dt_2.
    \end{multline}
\end{itemize}
As in the single-state case, an analytical solution can be obtained using Laplace techniques. The Laplace transform of the full polarisation function is
\begin{equation}
    F(s) = \int_0^{\infty} P(t) e^{-s t} dt = \sum_n \int_0^{\infty} R_n(t) e^{-s t} dt.
\end{equation}
This sum naturally separates into even and odd terms:
\begin{eqnarray}
    F(s) &=& \sum_n [ \nu^{2n} f_1^{n+1}(s) f_2^n(s) + \nu^{2n+1} f_1^{n+1}(s) f_2^{n+1}(s)]\cr
   & =& \frac{f_1(s) [1 + \nu f_2(s)]}{1 - \nu^2 f_1(s) f_2(s)},
    \label{eq:Lap2}
\end{eqnarray}
where $f_1(s)$ and $f_2(s)$ are the Laplace transforms of $P_{\rm stat1}(t)$ and $P_{\rm stat2}(t)$, respectively. This expression reduces to the known result (Eq.~\ref{eq:Lap}) when $f_1(s) = f_2(s)$. A more general form, allowing for asymmetric switching rates $\nu_1 \neq \nu_2$ and uncertainty in the initial state, yields
\begin{multline}
    F(s) = g \cdot \frac{f_1(s)\left[1 + \nu_1 f_2(s)\right]}{1 - \nu_1 \nu_2 f_1(s) f_2(s)} \\
    + (1 - g) \cdot \frac{f_2(s)\left[1 + \nu_2 f_1(s)\right]}{1 - \nu_1 \nu_2 f_1(s) f_2(s)},
    \label{eq:LapGen}
\end{multline}
where $g \in [0,1]$ is a weighting factor corresponding to the probability that the muon starts in state 1. Again, the single-state expression is recovered when $f_1(s) = f_2(s)$ and $\nu_1 = \nu_2$.

Since $\mu^+$SR experiments measure polarisation in the time domain, a direct solution for $P(t)$ is required for practical applications. While inverse Laplace transforms are in principle possible, they are computationally intensive. Fortunately, the structure of $R_n(t)$ allows a recursive formulation:
For even $n = 2\ell$ (with $\ell \in \mathbb{N}_0$):
\begin{equation}
    R_n(t) = \nu \int_0^t R_0(t_1) R_{n-1}(t - t_1) \, dt_1,
    \label{eq:recur2_1}
\end{equation}
and for odd $n = 2\ell + 1$:
\begin{equation}
    R_n(t) = \nu \int_0^t R_1(t_1) R_{n-2}(t - t_1) \, dt_1.
    \label{eq:recur2_2}
\end{equation}
The full dynamical polarisation function can then be expressed as:
\begin{multline}
    P_{\rm dyn}(t) = \sum_n R_n(t) = R_0(t) + R_1(t) \\
    + \nu \int_0^t \left[R_0(t_1) + R_1(t_1)\right] \sum_{n \geq 2} R_n(t - t_1) \, dt_1.
    \label{eq:recur2_3}
\end{multline}
By recursively applying Eq.~\ref{eq:recur2_3}, the polarisation function can be efficiently computed numerically. A key advantage of this formulation—shared with the nominal strong collision model—is that the dynamical function can be constructed directly from knowledge of the static functions, $P_{\rm stat1}(t)$ and $P_{\rm stat2}(t)$, without requiring detailed knowledge of the microscopic dynamics beyond the switching rate(s).

\section{\label{sec:result} Polarization Results}
To illustrate how the two-state model modifies the resulting muon polarisation, the polarisation function for two static functions is numerically computed: cosine (Eq.~\ref{eq:cos}) and Gaussian Kubo-Toyabe (G-KT; Eq.~\ref{eq:kt}) \cite{Kubo1967}. Specifically, these are defined as:
\begin{eqnarray}
P(t) = \cos(\gamma_\mu B t),
\label{eq:cos}
\end{eqnarray}
\begin{eqnarray}
P(t) = \frac{1}{3} + \frac{2}{3} (1 - \Delta^2 t^2 ) e^{ -\frac{1}{2} \Delta^2 t^2 }
\label{eq:kt}
\end{eqnarray}
where $\gamma_\mu$ is the gyromagnetic ratio of the muon, $B$ is the local magnetic field magnitude and $\Delta$ is the local magnetic field distribution width. The cosine function describes coherent precession typically found in magnetically ordered systems, whereas the G-KT function models isotropically distributed, randomly oriented magnetic moments \cite{Yaouanc2011}.

{\color{black}The two-state extension of Eq.~\ref{eq:cos} and Eq.~\ref{eq:kt} is shown in Fig.~\ref{fig:figure2}. The fluctuation rate $\nu$ is expressed in absolute units ($\mu$s$^{-1}$), corresponding to a correlation time $\tau_c = 1/\nu$. The dynamical regime is determined by comparing $\nu$ to the characteristic static field scale—either the field distribution width $\Delta$ or the precession frequency $\gamma_\mu B$. The quasi-static regime corresponds to $\nu \ll \Delta$, where the field changes slowly compared to the local field strength. Conversely, the motional narrowing limit is reached when $\nu \gg \Delta$, and rapid fluctuations average out the local fields.
}

When the muon are primarily couple to nuclear rather than electronic moments, the characteristic field width, $\Delta$, of the G-KT are typically below $0.4~\mu$s$^{-1}$ \cite{McClelland2020}. Applying the two-state model to this regime yields the behaviour shown in Fig.~\ref{fig:figure2}(a). In the static limit ($\nu = 0$), the resulting polarisation is simply the average of two static KT functions with different $\Delta$ values, producing a broader minimum in the polarisation curve. As $\nu$ increases, the lineshape evolves into a damped exponential, while still retaining a broadened $1/3$ tail and initially Gaussian form. In the motional narrowing limit ($\nu \gg \Delta$), the polarisation function becomes nearly exponential, resembling results from single-state model.

{\color{black}On the other hand, the field distribution widths can be much larger when muons couple to randomly oriented electronic moments, as in magnetically disordered systems such as short-range ordered phases or magnetic domains. The two-state model applied in this regime produces the behaviours shown in Fig.~\ref{fig:figure2}(b).} The characteristic 1/3 tail of the G-KT is suppressed by field fluctuations, though the preceding dip remains broader than in a standard KT shape in the quasi-dynamical regime. As the fluctuation rate $\nu$ increases further, the dip disappears entirely, and the lineshape becomes more Lorentzian-like—while still retaining a Gaussian onset. This intermediate regime closely resembles the phenomenological Gaussian-broadened Kubo-Toyabe function developed by Noakes and Kalvius (NK) \cite{Noakes1997}.

An important use of the two-state model lies in magnetically ordered systems, where the static polarisation typically follows a cosine form. This case is shown in Fig.~\ref{fig:figure2}(c,d), where the two state model is applied to static cosine function. At high fluctuation rates $\nu$, the polarisation becomes exponentially damped—both for equal-magnitude fields (Fig.~\ref{fig:figure2}(d)) and for unequal fields (Fig.~\ref{fig:figure2}(c)). Although such damping is often attributed to static field distributions, these results show that dynamical fluctuations alone can suppress coherent oscillations, highlighting the flexibility of the two-state formalism.

\section{\label{sec:Dis} Real systems}
The two-state collision model, as the name suggests, is applicable to systems in which the initial static field is altered by stochastic field fluctuations. Many real-world physical systems exhibit such behaviour. One example is charge order fluctuations in correlated electron systems, where valence transitions can alter the magnetic properties of an ion \cite{Naveed2011, Bhobe2015, Khasanov2022}. For instance, a site may stochastically switch between different charge (and therefore spin) states, leading to distinct local magnetic environments. If these fluctuations occur between two dominant valence states with well-defined field signatures, the two-state model provides a natural framework for describing the resulting depolarization.

Other relevant examples include dynamic processes in superparamagnetic nanoparticles, domain wall motion in ferromagnets, and spin dynamics in Ising-like magnets, all of which involve discrete or quasi-discrete changes in the local magnetic field. In such systems, each fluctuation can be seen as a "collision" that transitions the muon from one static field configuration to another, making the two-state formalism particularly effective in capturing their dynamic behaviour. These fluctuations violates the core assumption of the nominal strong collision model—namely, that each fluctuation resets the system to a state described by the same static polarisation function (as given in Eq.~\ref{eq:Lap}).

\begin{figure}[ht]
  \begin{center}
    \includegraphics[keepaspectratio=true,width=75 mm]{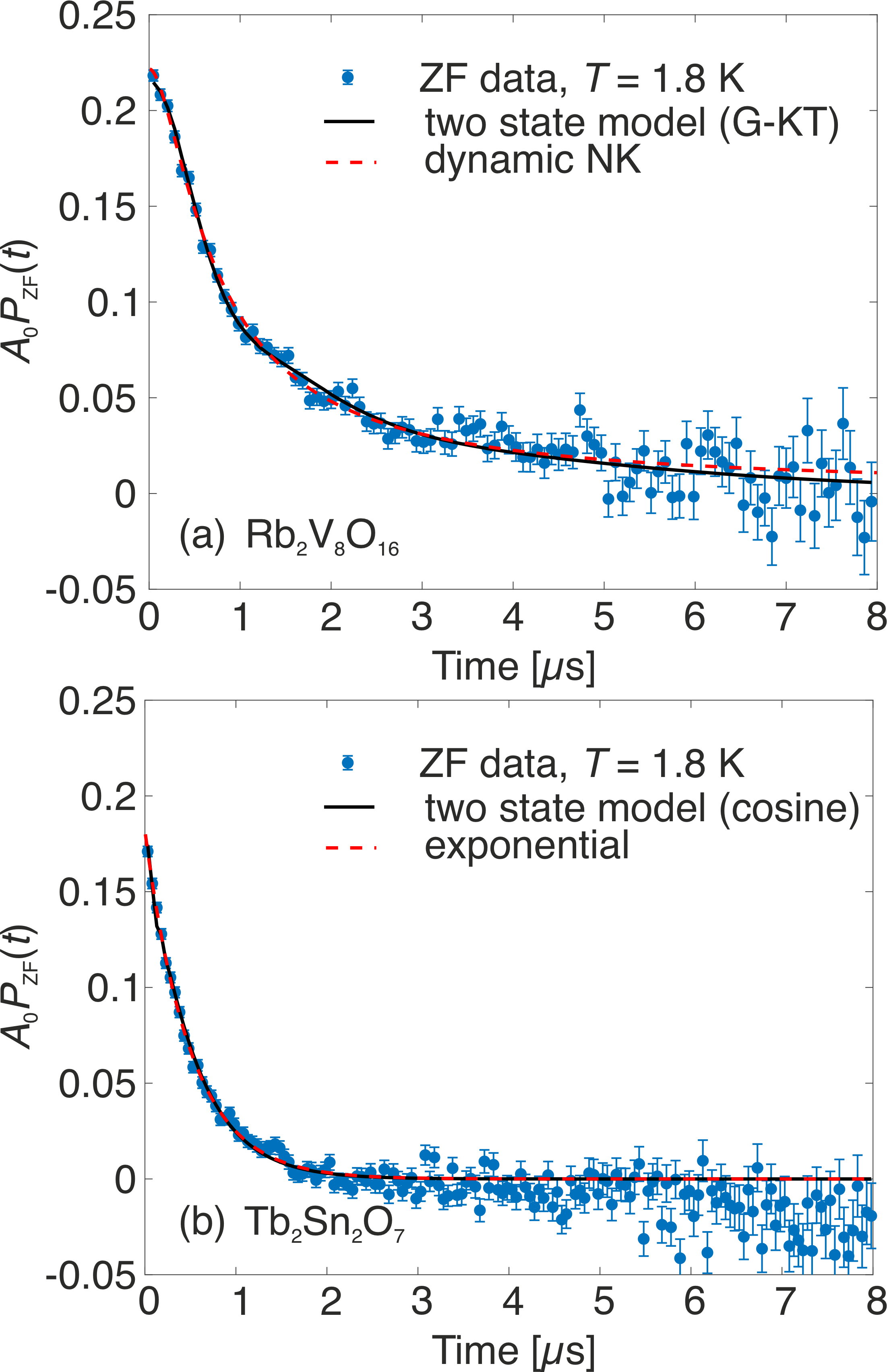}
  \end{center}
  \caption{{\color{black}
(a) ZF $\mu^+$SR time spectrum (scattered points) of Rb$_2$V$_8$O$_{16}$ at 1.8~K, reproduced from Ref.~\onlinecite{Forslund2021}. The solid line shows a fit using the two-state model based on the Gaussian Kubo-Toyabe (G-KT) function (Eq.~\ref{eq:kt}), yielding the parameters $A = 0.214(6)$, $\nu = 0.560(65)$~$\mu$s$^{-1}$, $\Delta_1 = 1.716(178)$~$\mu$s$^{-1}$, $\Delta_2 = 0.491(37)$~$\mu$s$^{-1}$, and $g = 0.535(78)$. The dashed line shows a fit using the dynamic Noakes-Kalvius (NK) function \cite{Noakes1997}. (b) ZF $\mu^+$SR time spectrum (scattered points) of Tb$_2$Sn$_2$O$_7$ at 1.8~K, from Ref.~\onlinecite{Dalmas2006}. The solid line shows a fit using the two-state model based on the cosine function (Eq.~\ref{eq:cos}), with fit parameters $A = 0.182(15)$, $\nu = 38.91(8.68)$~$\mu$s$^{-1}$ and $B = 16.3(2.9)$~mT. In this fit approach, $g = 0.5$ and the local fields $B_2 = -B_1$ were fixed. The dashed line shows a simple exponential fit for comparison.
}
}

  \label{fig:figure3}
\end{figure} 

To demonstrate the applicability of the two-state generalisation, {\color{black} a zero field (ZF) $\mu^+$SR spectrum on the mixed-valence vanadium oxide compound Rb$_2$V$_8$O$_{16}$} -- previously analysed using the NK function~\cite{Forslund2021}-- is presented in Fig.~\ref{fig:figure3}(a). {\color{black}Rb$_2$V$_8$O$_{16}$ is a quasi-1D compound, undergoing a metal–insulator transition near 200~K accompanied by charge ordering of V$^{3+}$ and V$^{4+}$ \cite{Isobe2009}. Below this transition, magnetic susceptibility shows a drop followed by a Curie–Weiss tail at lower temperatures, consistent with residual fluctuating moments. }

As seen, the two-state model based on the G-KT static functions provides an equally good fit and is, in practice, indistinguishable from the NK function. However, the microscopic interpretations of the two approaches differ. The NK function is largely phenomenological, though supported by Monte Carlo simulations \cite{Noakes1999} in systems with spatially modulated short-range order or disorder. The two-state model, on the other hand, implies a dynamic process in which each fluctuation leads to a distinct change in the underlying static polarisation function. These results demonstrates that NK-like depolarisation lineshapes can arise not only from spatial disorder or short-range correlations but also from dynamical systems where the local field environment changes between well-defined states (the two state generalisation of the strong collision model). As such, the two-state collision model provides a complementary — and in some cases, more physically transparent—framework for interpreting $\mu^+$SR spectra in complex, fluctuating systems.

{\color{black}The two-state model interpretation directly links the observed depolarisation to physically meaningful parameters such as the fluctuation rate $\nu \approx 0.56~\mu s^{-1}$ and field widths $\Delta_1 \approx 1.7~\mu s^{-1}$ and $\Delta_2 \approx 0.5~\mu s^{-1}$. A natural interpretation is that the muon senses the dynamic motion of magnetic domains or fluctuating short-range magnetic order. As such domains move past the muon site, the local magnetic environment stochastically switches between regions of stronger and weaker magnetization, producing the observed two-state behaviour. Alternatively, the switching between field distributions may reflect valence fluctuations between V$^{3+}$ and V$^{4+}$ ions, corresponding to different local spin states. Such fluctuations have been observed in the related quasi-1D mixed-valence compound K$_2$Cr$_8$O$_{16}$~\cite{Bhobe2015}. The near-equal population parameter ($g$) indicates the muons sample these environments approximately equally, consistent with a microscopically heterogeneous ground state with coexisting phases or dynamic spatial inhomogeneity.
}

{\color{black}Another important application of the two-state collision model is in fluctuating magnetically ordered systems, such as the pyrochlore compound Tb$_2$Sn$_2$O$_7$~\cite{Dalmas2006}.} While neutron diffraction measurements reveals magnetic Bragg peaks, confirming long-range magnetic order, $\mu^+$SR measurements show a purely exponential depolarisation. To reconcile this discrepancy, the local field was assumed to fluctuates between two opposite directions with equal magnitude. {\color{black}Drawing from studies on similar compounds, such as Gd$_2$Ti$_2$O$_7$ \cite{Yaouanc2005} and Gd$_2$Sn$_2$O$_7$ \cite{Bonville2004}, a local field of approximately 200~mT was assumed. Under these conditions, it was shown that the system satisfies the motional narrowing criterion. While this explanation is viable, it depends on a specific symmetry in the local field dynamics.} 

In contrast, the two-state collision model offers a more natural and general description of this behaviour, without requiring specific assumptions about the field symmetry. {\color{black}Figure~\ref{fig:figure3}(b) shows a fit of the ZF time spectrum of Tb$_2$Sn$_2$O$_7$ reported in Ref.~\onlinecite{Dalmas2006}, using the cosine-based two-state collision model with a similar symmetry constraint imposed for comparison. The fit yields a local field of $B = 16.3(2.9)$~mT and a fluctuation rate of $\nu = 38.9(8.7)$~$\mu$s$^{-1}$. Although the motional narrowing condition is fulfilled here as well ($\gamma_\mu B / \nu \approx 0.06 \ll 1$), a key difference from the analysis in Ref.~\onlinecite{Dalmas2006} is the extracted local field, which is an order of magnitude smaller than the previously assumed 200~mT. 

To assess the internal field without relying on model-specific assumptions, consider the longitudinal-field scan shown in Fig.~3 of Ref.~\cite{Dalmas2006}. While the field dependence does not strictly follow Redfield behaviour, the data show that decoupling begins around $15-20$~mT at 1.8~K. This supports the possibility that the true local field is indeed closer to the extracted value of $\sim16$~mT, rather than the previously assumed 200~mT. These results show that the two-state model can yield physically meaningful parameters even deep within the motional narrowing limit, where the signal reduces to a simple exponential -- even though its primary strength lies in describing the quasi-dynamic regime, where oscillatory structure is retained in the spectrum (Fig.~\ref{fig:figure3}(c,d)). The fact that the local field magnitude is an order of magnitude smaller than previously expected suggests that spin dynamics in pyrochlores may involve mechanisms which effectively reduce the local field -- mechanisms that warrant further investigation.
}

{\color{black}Finally, although not demonstrated in this work, it is worth discussing how} the two-state model {\color{black}may} find application in the study of ion diffusion, particularly in battery materials such as LiCoO$_2$ and NaCoO$_2$ \cite{Sugiyama2009, Maansson2013}. In these systems, the diffusion of Li or Na ions leads to site-to-site migration—sometimes via interstitial sites—which alters the nuclear dipolar field experienced by the muon \cite{Borg2012}. As the local magnetic environment fluctuates with the ionic configuration, the assumptions underlying Eq.~\ref{eq:Lap} may no longer hold. Depending on whether the ions hop between magnetically identical or non-identical environments, the appropriate description may shift between Eq.~\ref{eq:Lap} and Eq.~\ref{eq:LapGen}.

{\color{black} More broadly, the two-state model may offer new avenues for addressing a long-standing challenge in $\mu^+$SR: distinguishing between local magnetic field fluctuations and muon diffusion processes \cite{Sugiyama2020}. These two mechanisms affect the muon’s magnetic environment in fundamentally different ways. Static muons subjected to intrinsic magnetic fluctuations may or may not exhibit distinct switching behaviour, depending on the nature of the system. Muon diffusion, by contrast, involves hopping between crystallographic or interstitial sites and naturally gives rise to stochastic switching between distinct local magnetic environments. This motion breaks the assumption that the polarisation function remains unchanged between transitions. Instead, the two-state model becomes particularly relevant, as it directly captures the dynamics of transitions between discrete magnetic environments, offering a physically motivated framework that may help disentangle these two mechanisms in appropriate systems.}

\section{\label{sec:Con} Conclusion}
In this work, the strong collision model has been extended to address key limitations in the treatment of dynamical effects. While the nominal strong collision model provides a foundational framework for incorporating stochastic fluctuations in the local magnetic field, it assumes that each fluctuation returns the system to the same static environment. The two-state model introduced here allows for state changes between distinct local environments during each collision, offering a more flexible and realistic description of polarisation behaviour in such cases. By solving the model both analytically and numerically, its applicability to a range of condensed matter systems is demonstrated. While the two-state generalisation is presented from a $\mu^+$SR perspective in this work, the underlying stochastic field-switching concept can, in principle, be applied to a broader class of relaxation phenomena. These results provide an improved theoretical basis for analysing data in complex materials, potentially broadening the scope and interpretive power of the technique across a variety of physical and chemical systems.

\begin{acknowledgments}
I am supported by the Swedish Research Council (VR) through Grant 2022-06217, the Foundation Blanceflor fellowships for 2023 and 2024, and the Ruth and Nils-Erik Stenbäck Foundation. I would like to thank my daughter, Elisa Astrid Kuno Forslund, for her constant energy, which has been a source of both motivation and inspiration throughout this work.

\end{acknowledgments}

\section{Data Availability}
{\color{black}The data are available from the author upon reasonable request.}

\bibliography{Refs} 
\end{document}